\documentstyle[12pt]{article}
\begin{document}
\title{A QCD Generated Mass Spectrum}
\author{B.G. Sidharth\\
B.M. Birla Science Centre, Adarshnagar, Hyderabad - 500 063,
India}
\date{}
\maketitle
\begin{abstract}
Using the interquark potential we obtain a formula for the mass spectrum of elementary particles. The simple formula gives the masses of all known elementary particles with an error of about three percent or less. This includes the recently discovered $Ds (2317)$ and $1.5GeV$ Pentaquark particles.
\end{abstract}
One of the problems that has eluded a solution is that of a mass spectrum for elementary particles. Such a formula would be intimately tied up with inter quark interactions. We will now use the QCD potential to deduce such a formula, which as will be seen, surprisingly covers all known elementary particles. The well known QCD potential is given by \cite{lee,cu}
\begin{equation}
U(r) = - \frac{\alpha}{r} + \beta r\label{e1}
\end{equation}
where in units $\hbar = c = 1, \alpha \sim 1$. The first term in (\ref{e1}) represents the Coulumbic part while the second term represents the confining part of the potential.\\

Let us consider the pion made up of two quarks along with a third quark, one at the centre and two at the ends of interval of the order of the Compton wavelength, $r$. Then the central particle experiences the force
\begin{equation}
\frac{\alpha}{(\frac{r}{2} + 2)^2} - \frac{\alpha}{(\frac{r}{2} -
x)^2} \approx \frac{-2\alpha x}{r^3}\label{e2}
\end{equation}
where $x$ is the small displacement from the mean position. Equation (\ref{e2}) gives rise to the Harmonic Oscillator potential, and the whole configuration resembles the tri-atomic molecule.\\
Before proceeding we can make a quick check on (\ref{e2}). We use the fact that the frequency is given by
$$\omega = \left(\frac{\alpha^2}{m_\pi r^3}\right)^{\frac{1}{2}} = \frac{\alpha}{(m_\pi r^3)^{\frac{1}{2}}}$$
whence the mass of the pion $m_\pi$ is given by
\begin{equation}
(h \omega \equiv ) \omega = m_\pi\label{e3}
\end{equation}
Remembering that $r = 1/m_\pi$, use of (\ref{e3}) gives $\alpha = 1$, which ofcourse is correct.\\
To proceed the energy levels of the Harmonic oscillator are now given by,
$$\left(n + \frac{1}{2}\right) m_\pi ,$$
or if there are small $m$ such oscillators, we have
\begin{equation}
E = m_P  = m \left(n + \frac{1}{2}\right) m_\pi\label{e4}
\end{equation}
where $m_P$ is the mass of the corresponding elementary particle. The formula (\ref{e4}) gives the mass of all known elementary particles with an error of less than one percent for sixty three percent of the particles, less than two percent for ninety three percent of the particles, and less than three percent for all particles with the lone exception of $\omega (782)$, in which case the error is $3.6 \%$. The known elementary particles for which the formula (\ref{e4}) is valid include the recently discovered $Ds(2317)$ and the $1.5 GeV$ Pentaquark.\\ \\
{\large \bf{Remarks:}}\\ \\
Firstly it is surprising that there is such a good fit for all the particles \cite{hagiwara} considering that only bare details of the interaction have been taken into consideration. Once other details are included, the agreement could be even better.  Secondly, it may be mentioned that a similar approach, but using the proton as the base particle had lead to interesting, but not such comprehensive results \cite{mpla1,mpla2,lobanov,csf,ffp5}.

\newpage
\begin{table}
\caption{Baryons}
\begin{tabular}{|c|c|c|c|} \hline
Particle and Mass & Mass from Formula &  Error \% & $(m,n)$ \\ \hline
   $p(938)$ & $959$ &  $-2.23881$, & $(2,3)$ \\
   $n(939)$ & $959$ &  $-2.12993,$ & $(2,3)$ \\
$P_{11} **** N(1440)$  & $1438.5$ & $(0.138889,)0$ & $(1,10)$ \\
$D_{13}**** N(1520)$ & $1507$ &  $(0.855263,)$ & $(2,5)$\\
$S_{11}****N(1535)$ & $1507$  & $1.9442$ & $(2,5)$ \\
$S_{11}****N(1650)$ & $1644$  & $(0.363636,)0$ & $(8,1)$ \\
$D_{15}****N(1675)$ & $1644$  & $1.85075,$ & $(8,1)$ \\
$F_{15}****N(1680)$ & $1644$  & $2.14286,$ & $(8,1)$ \\
$D_{13}***N(1700)$  & $1712.5$  & $(-0.705882,)0$ & $(1,12)$ \\
$P_{11}***N(1710)$ & $1712.5$   & $(-0.116959,)0$ & $(1,12)$ \\
$P_{13}****N(1720)$ & $1712.5$  & $(0.465116,)0$ & $(1,12)$\\
$P_{13}****N(1900)$ & $1918$  & $-0.947368,$ & $(4,3)$ \\
$F_{17}**N(1990)$ & $1986.5$    & $0.201005,$ & $(1,14)$\\
$F-{15}**N(2000)$ & $1986.5$    & $0.7,$  & $(1,14)$\\
$D_{13}**N(2080)$ & $2055$    & $1.20192,$ & $(2,7)$ \\
$S_{11}*N(2090)$ & $2123.5$     & $-1.57895,$ & $(1,15)$ \\
$P_{11}*N(2100)$ & $2123.5$     & $(-1.09524,)$ & $(1,15)$ \\
$G_{17}****N(2190)$ & $2123.5$  & $(3.05936,)0$  & $(1,15)$\\
$D_{15}**N(2200)$ & $2260.5$  & $-2.72727,$ & $(3,5)$ \\
$H_{19}****N(2220)$ & $2260.5$  & $(-1.8018,)0$ & $(3,5)$ \\
$G_{19}****N(2250)$ & $2260.5$  & $(-0.444444,)0$ & $(3,5)$ \\
$I_{1;11}***N(2600)$ & $2603$ & $(-0.115385,)0$ & $(2,9)$ \\
$K_{1;13}**N(2700)$ & $2671.5$ & $1.05556$ & $(1,19)$\\
$P_{33}****\Delta (1232)$ & $1233$ & $(-0.0811688,)0$ & $(2,4)$ \\
$P_{33}***\Delta (1600)$ & $1575.5$  & $(1.5625,)0$ & $(1,11)$ \\
$S_{31}****\Delta (1620)$ & $1644$ & $(-1.46148,)0$ & $(8,1)$ \\
$D 33****\Delta (1700)$ & $1712$  & $(-0.705882,)0$ & $(1,12)$ \\
$P 31*\Delta (1750)$ & $1781$  & $-1.77143,$ & $(2,6)$ \\
$S_{31}**\Delta (1900)$ & $1918$  & $-0.947368,$ & $(4,3)$ \\
$F 35****\Delta (1905)$ & $1918$  & $(-0.682415,)0$ & $(4,3)$\\
$P 31****\Delta (1910)$ & $1918$ & $(-0.418848,)0$ & $(4,3)$ \\
$P 33***\Delta (1920)$ & $1918$  & $(0.104167,)0$ & $(4,3)$ \\
$D 35***\Delta (1930)$ & $1918$  & $(0.621762,)0$ & $(4,3)$\\
$D 33*\Delta (1940)$ & $1918$  & $1.13402,$ & $(4,3)$ \\ \hline
\end{tabular}
\end{table}

\newpage

\begin{table}
\begin{tabular}{|c|c|c|c|} \hline
Particle and Mass & Mass from Formula &  Error \%  & $(m,n)$\\ \hline
$F 37****\Delta (1950)$ & $1918$  & $1.64103,$ & $(4,3)$ \\
$F 35**\Delta (2000)$ & $1986$  & $0.7,$ & $(1,14)$\\
$S_{31}*\Delta (2150)$ & $2123.5$ & $1.25581,$ & $(1,15)$ \\
$G_{37}*\Delta (2200)$ & $2260$ & $-2.72727,$ & $(1,16)$ \\
$H_{39}**\Delta (2300)$ & $2329$  & $-1.26087,$ & $(2,8)$ \\
$D_{35}*\Delta (2350)$ & $2329$  & $0.893617,$ & $(2,8)$\\
$F_{37}*\Delta (2390)$ & $2397.5$  & $-0.292887,$ & $(1,17)$ \\
$G_{39}**\Delta (2400)$ & $2397.5$ & $0.125,$ & $(1,17)$ \\
$H_{3;11}****\Delta (2420)$ & $2397.5$  & $(0.950413,)0$ & $(1,17)$ \\
$I_{3;13}**\Delta (2750)$ & $2740$  & $0.363636,$ & $(8,2)$\\
$K_{3;15}**\Delta (2950)$ & $2945.5$  & $0.152542,$ & $(1,21)$ \\
              $\Lambda (1115)$ & $1096$  & $1.7000,$ & $(16,0)$ \\
$P_{01}****\Lambda (1600)$  & $1575.5$  &  $1.53125,$ & $(1,11)$ \\
$S_{01}****\Lambda (1405)$ & $1438.5$ &  $-2.3130,$ & $(1,10)$ \\
$D_{03}****\Lambda (1520)$ & $1507$ &  $0.855263,$ & $(2,5)$\\
$P 01***\Lambda (1600)$ & $1575.5$ &  $(1.5625,)0$ & $(1,12)$\\
$S 01****\Lambda (1670)$ & $1644$  &  $1.55689,$ & $(8,1)$ \\
$D 03****\Lambda (1690)$ & $1712.5$  &  $-1.30178,$ & $(1.12)$ \\
$S 01***\Lambda (1800)$ & $1781$  & $(1,05556,)0$ & $(2,6)$ \\
$P 01***\Lambda (1810)$ & $1781$  & $(1.60221,)0$ & $(2,6)$ \\
 $\Lambda (1820)$ & $1849.5$ & $(2.14286,)$ & $(1,13)$ \\
$D 05****\Lambda (1830)$ & $1849.5$ & $-1.03825,$ & $(1,13)$ \\
$P 03****\Lambda (1890)$ & $1918$  & $-1.48148,$ & $(4,3)$\\
$*\Lambda (2000)$ & $1986.5$  & $0.7,$ & $(1,14)$  \\
$F 07*\Lambda (2020)$ & $2055$ & $-1.73267,$ & $(2,7)$\\
$G 07****\Lambda (2100)$ & $2123.5$  & $-1.09524,$ & $(1,15)$ \\
$F 05***\Lambda (2110)$ & $2123.5$ & $(-0.616114,)0$ & $(1,15)$ \\
$D 03*\Lambda (2325)$ & $2329$  & $-0.172043,$ & $(2,8)$\\
$H 09***\Lambda (2350)$ & $2329$  & $0.893617,$ & $(2,8)$ \\
$**\Lambda (2585)$ & $2603$ & $0.309478,$ & $(2,9)$ \\
$P 11****\Sigma +(118)$  & $1164.5$ & $2.10261,$ & $(1,8)$ \\
$P 11****\Sigma 0 (119)$ & $1164.5$ & $2.34899,$ & $(1,8)$ \\
$****\Sigma - (119)$ & $1164.5$  & $2.75689,$ & $(1,8)$ \\
$P 13****\Sigma (1385)$  & $1370$ & $(0.108),$ & $(4,2)$ \\
$*\Sigma (1480)$  & $1438.5$ & $2.83784,$ & $(1,10)$ \\
$**\Sigma (1560)$  & $1575.5$   & $-0.961538,$ & $(1,11)$\\ \hline
\end{tabular}
\end{table}

\newpage

\begin{table}
\begin{tabular}{|c|c|c|c|} \hline
Particle and Mass & Mass from Formula  & Error \% & $(m,n)$ \\ \hline
$D 13**\Sigma (1580)$  & $1575.5$ & $0.316456,$ & $(1,11)$ \\
$S 11**\Sigma (1620)$  & $1644$ & $-1.48148,$ & $(8,1)$ \\
$P 11***\Sigma (1660)$  & $1644$ & $(0.963855,)0$ & $(8,1)$ \\
$D 13****\Sigma (1670)$  & $1644$ & $1.55689,$ & $(8,1)$ \\
$**\Sigma (1690)$  & $1712.5$  & $-1.30178,$ & $(1,12)$   \\
$S 11***\Sigma (1750)$  & $1781$ & $(-1.77143,)0$ & $(2,6)$ \\
$P 11*\Sigma (1770)$  & $1781$  & $-0.621469,$ & $(2,6)$\\
$D 15****\Sigma (1775)$  & $1781$ & $(-0.338028,)0$ & $(2,6)$ \\
$P 13*\Sigma (1840)$  & $1849.5$ & $-0.48913,$ & $(1,13)$ \\
$P 11**\Sigma (1880)$  & $1849.5$   & $1.64894,$ & $(1,13)$ \\
$F 15****\Sigma (1915)$  & $1918$ & $(-0.156658,)0$ & $(4,3)$ \\
$D 13***\Sigma (1940)$  & $1918$ & $(1.13402,)0$ & $(4,3)$ \\
$S 11*\Sigma (2000)$  & $1986.5$ & $0.7,$ & $(1,14)$  \\
$F 17****\Sigma (2030)$  & $2055$ & $-1.23153,$ & $(2,7)$ \\
$F 15*\Sigma (2070)$  & $2055$ & $0.724638,$ & $(2,7)$ \\
$P 13**\Sigma (2080)$  & $2055$ & $1.20192,$ & $(2,7)$ \\
$G 17*\Sigma (2100)$  & $2123$ & $-1.09524,$ & $(1,15)$ \\
$***\Sigma (2250)$  & $2260$   & $(-0.444444,)0$ & $(3,5)$ \\
$**\Sigma (2455)$  & $2466$ & $-0.448065,$ & $(4,4)$ \\
$**\Sigma (2620)$  & $2603$ &  $0.648855,$ & $(2,9)$ \\
$*\Sigma (3000)$  & $3014$  & $-0.466667,$ & $(4,5)$ \\
$*\Sigma (3170)$  & $3151$  & $0.599369,$ & $(2,11)$ \\
$P 11****\Xi 0,\Xi-(13)$ & $1301.5$  & $1.01156,$ & $(1,9)$ \\
$****\Xi(1321)$  & $1301.5$  & $1.47615,$ & $(1,9)$ \\
$P 13****\Xi (1530)$     & $1507$  & $1.50327,$ & $(2,5)$ \\
$*\Xi (1620)$  & $1644$  & $-1.48148,$ & $(8,1)$ \\
$***\Xi (1690)$     & $1712.5$  & $-1.30178,$ & $(1,12)$ \\
$D 13***\Xi (1820)$     & $1849.5$ & $-1.59341,$ & $(1,13)$ \\
$***\Xi (1950)$   & $1918$   & $1.64103,$ & $(4,3)$  \\
$***\Xi (2030)$     & $2055$ & $-1.23153,$ & $(2,7)$ \\
$*\Xi (2120)$     & $2123.5$   & $-0.141509,$ & $(1,15)$ \\
$**\Xi (2250)$     & $2260.5$  & $-0.444444,$ & $(1,16)$ \\
$**\Xi (2370)$     & $2397.5$  & $-1.13924,$ & $(1,17)$ \\
$*\Xi (2500)$     & $2534.5$   & $-1.36,$ & $(1,18)$   \\ \hline
\end{tabular}
\end{table}
\newpage

\begin{table}
\begin{tabular}{|c|c|c|c|} \hline
Particle and Mass & Mass from Formula  & Error \% & $(m,n)$ \\ \hline
$****\Omega -(1672)$ & $1644$  & $1.67464,$ & $(8,1)$ \\
$***\Omega -(2250)$ & $2260.5$   & $(-0.444444,)0$ & $(1,16)$\\
$**\Omega -(2380)$ & $2397.5$   & $-0.714286,$ & $(1,17)$ \\
$**\Omega -(2470)$ & $2466$  & $0.161943,$ & $(4,4)$ \\
$****\Lambda c+2285)$  & $2260.5$ & $1.09409,$ & $(1.16)$ \\
$+***\Lambda c+(2593)$ & $2603$ & $-0.385654,$ & $(2,9)$ \\
$+***\Lambda c+(2625)$ & $2603$ & $0.838095,$ & $(2,9)$\\
$+*\Lambda c+(2765)$ & $2740$  & $0.904159,$ & $(8,2)$\\
$+**\Lambda c+(2880)$ & $2877$  & $0.104167,$ & $(2,10)$ \\
$****\Sigma c(2455)$   & $2466$  & $-0.448065,$ & $(4,4)$ \\
$***\Sigma c(2520)$   & $2534.5$  & $-0.555556,$ & $(1,18)$ \\
$\Xi c+(2466)$   & $2466$  & $0,$  & $(4,4)$        \\
$***\Xi c0(2471)$     & $2466$  & $0.202347,$ & $(4,4)$  \\
$***\Xi c+(2574)$     & $2603$  & $(1.12665,)0$ & $(2,9)$ \\
$***\Xi c0(2578)$     & $2603$  & $(0.96974,)0$ & $(2,9)$ \\
$\Xi c(2645)$      & $2671.5$     & $-0.982987,$ & $(1,19)$\\
$***\Xi c(2790)$      & $2808.5$  & $-0.645161,$ & $(1,20)$\\
$***\Xi c(2815)$      & $2808.5$  & $0,248668,$ & $(1,20)$\\
$***\Omega c0(2697)$  & $2671.5$  & $0.964034,$ & $(1,19)$\\
$***\Lambda b0(5624)$ & $5617$  & $(0.124467,)0$ & $(2,20)$ \\ \hline
\end{tabular}
\end{table}

\newpage

\begin{table}
\caption{Mesons}
\begin{tabular}{|c|c|c|c|} \hline
Particle and mass & Mass From Formula &  Error \% & $(m,n)$\\ \hline
$*\pi^{\pm} (139)$ &  $137$ & $-1.43885$ \\
$*\pi^0 (135)$ & $137$ & $1.481481$  \\
$*\eta (547)$ & $548$ &  $0.182815$ & $(8,0)$\\
$*f_0(600)$ & $616.5$ & $(2.75)0$ & $(1,4)$ \\
$*\rho (770)$ & $753.5$ & $-2.14286$ & $(1.5)$\\
$*\omega (782)$ & $753.5$ & $-3.6445$ & $(1,5)$\\
$*\eta' (958)$ & $959$ & $0.104384$ & $(2,3)$\\
$*f_0(980)$ & $ 959$ & $-2.14286$ & $(2,3)$\\
$*a_0(980)$ & $ 959$ & $-2.14286$ & $(2,3)$\\
$*\phi (1020)$ & $1027.5$ & $0.735294$ & $(1,7)$\\
$*h_1 (1170)$ & $1164.5$ & $(-0.47009)0$ & $(1,8)$\\
$*b_1 (1235)$ & $1233$ & $(-0.16194)0$ & $(2,4)$ \\
$a_1 (1260)$ & $1233$ & $(-2.14286)0$ & $(2,4)$ \\
$f_2 (1270)$ & $1233$ &  $-2.91339$ & $(2,4)$\\
$f_1 (1285)$ & $1301.5$ & $1.284047$ & $(1,9)$\\
$*\eta (1295)$ & $1301.5$ &  $0.501931$ & $(1,9)$\\
$\pi (1300)$ & $1301.5$ &  $0.115385$ & $(1,9)$\\
$a_2 (1320)$ & $1301.5$ & $-1.40152$ & $(1,9)$\\
$*f_0 (1370)$ & $1370$ & $0$ & $(4,2)$\\
$h_1 (1380)$ & $1370$ &  $0.72464$ & $(4,2)$\\
$\pi_1 (1400)$ & $1370$ & $-2.14286$ & $(4,2)$ \\
$f_1 (1420)$ & $1438.5$ & $1.302817$ & $(1,10)$\\
$*\omega (1420)$ & $1438.5$ & $(1.302817)0$ & $(1,10)$ \\
$f_2 (1430)$ & $1438.5$ & $0.594406$ & $(1,10)$ \\
$*\eta (1440)$ & $1438.5$ &  $-0.10417$ & $(1,10)$\\
$*a_0 (1450)$ & $1438.5$ &  $-0.7931$ & $(1,10)$\\
$*\rho (1450)$ & $1438.5$ & $-0.7931$ & $(1,10)$ \\
$*f_0 (1500)$ & $1507$ & $(0.466667)0$ & $(2,5)$ \\
$f_1 (1510)$ & $1507$ &  $-0.19868$ & $(2,5)$\\
$*f'_2 (1525)$ & $1507$ & $-1.18033$ & $(2,5)$\\
$f_2 (1565)$ & $1575.5$ & $0.670927$ & $(1,11)$ \\ \hline
\end{tabular}
\end{table}
\newpage

\begin{table}
\begin{tabular}{|c|c|c|c|} \hline
Particle and mass & Mass From Formula &  Error \% & $(m,n)$\\ \hline
$h_1 (1595)$ & $1575.5$ &  $-1.22257$ & $(1,11)$\\
$\pi_1 (1600)$ & $1575.5$ & $-1.53125$ & $(1,11)$ \\
$\chi (1600)$ & $1575.5$ & $-1.53125$ & $(1,11)$ \\
$a_1 (1640)$ & $1644$ & $0.243902$ & $(8,1)$ \\
$f_2 (1640)$ & $1644$ &  $0.243902$  & $(8,1)$ \\
$\eta_2 (1645)$ & $1644$ & $(0.06079)0$ & $(8,1)$ \\
$\omega (1670)$ & $1644$ & $(1.55688)0$ & $(8,1)$ \\
$*\omega_3 (1670)$ & $1644$ & $-1.55689$ & $(8,1)$ \\
$*\pi_2  (1670)$ & $1644$ & $-1.55689$ & $(8,1)$ \\
$*\phi (1680)$ & $1712.5$ & $1.934524$ & $(1,12)$ \\
$*\rho_3 (1690)$ & $1712.5$ & $1.331361$ & $(1,12)$ \\
$*\rho (1700)$ & $1712.5$ & $(0.735294)0$ & $(1,12)$ \\
$a_2 (1700)$ & $1712.5$ &  $0.735294$ & $(1,12)$ \\
$f_0(1710)$ & $1712.5$ & $(0.146199)0$ & $(1,12)$ \\
$\eta (1760)$ & $1781$ &  $1.193182$ & $(2,6)$ \\
$*\pi (1800)$ & $ 1781$ & $-1.05556$ & $(2,6)$ \\
$f_2(1810)$ & $1781$ & $-1.60221$ & $(2,6)$ \\
$*\phi_3 (1850)$ & $1849.5$ & $(-0.02703)0$ & $(1,13)$\\
$\eta_2 (1870)$ & $1849.5$ & $-1.09626$ & $(1,13)$ \\
$\rho (1900)$ & $1918$ &  $0.947368$ & $(4,3)$\\
$f_2 (1910)$ & $1918$ & $0.418848$ & $(4,3)$\\
$f_2 (1950)$ & $1918$ & $-1.64103$ & $(4,3)$\\
$\rho_3 (1990)$ & $1986.5$ & $-0.17588$ & $(1,14)$\\
$X (2000)$ & $1986.5$ & $-0.675$ & $(1,14)$\\
$f_2 (2010)$ & $1986.5$ & $(-1.16915)0$ & $(1,14)$ \\
$f_0 (2020)$ & $1986.5$ & $1.65842$ & $(1,14)$\\
$*a_4 (2040)$ & $2055$ & $0.735294$ & $(2,7)$\\
$f_4 (2050)$ & $2055$ &  $0.243902$ & $(2,7)$\\
$\pi_2 (2100)$ & $2123.5$ & $1.119048$ & $(1,15)$\\
$f_0 (2100)$ & $2123.5$ &  $1.119048$ & $(1,15)$\\
$f_2 (2150)$ & $2123.5$ &  $-1.23256$ & $(1,15)$\\ \hline
\end{tabular}
\end{table}

\newpage

\begin{table}
\begin{tabular}{|c|c|c|c|} \hline
Particle and mass & Mass From Formula &  Error \% & $(m,n)$\\ \hline
$\rho_2 (2150)$ & $2123.5$ & $-1.23256$ & $(1,15)$ \\
$f_0 (2200)$ & $2260.5$ & $2.75$ & $(1,16)$\\
$f_J (2220)$ & $2260.5$ &  $1.824324$ & $(1,16)$\\
$\eta (2225)$ & $2360$ & $1.595506$ & $(1,16)$\\
$\rho_3 (2250)$ & $2260$ & $0.466667$ & $(1,16)$\\
$*f_2 (2300)$ & $2329$ & $1.26087$ & $(2,8)$ \\
$f_4 (2300)$ & $2329$ &  $1.26087$ & $(2,8)$\\
$D_s (2317)$ & $2329$ & $0.5$  &$(2,8)$\\
$f_0 (2330)$ & $2329$ &  $-0.04292$ & $(2,8)$ \\
$*f_2 (2340)$ & $2329$ & $-0.47009$ & $(2,8)$ \\
$\rho_5 (2350)$ & $2329$ & $-0.89362$ & $(2,8)$ \\
$a_6 (2450)$ & $2466$ & $-0.89362$ & $(4,4)$ \\
$f_6 (2510)$ & $2534.5$ & $0.976096$ & $(1,18)$ \\
$*K^* (892)$ & $890.5$ & $-0.16816$ & $(1,6)$\\
$*K_1(1270)$ & $1233$ & $2.91338$ & $(2,4)$ \\
$*K_1(1400)$ & $1370$ & $-2.14286$ & $(4,2)$\\
$*K^*(1410)$ & $1438.5$ & $2.021277$ & $(1,10)$  \\
$*K^*_0(1430)$ & $1438.5$ & $0.594406$ & $(1,10)$ \\
$*K^*_2(1430)$ & $1438.5$ & $0.594406$ & $(1,10)$ \\
$K (1460)$ & $1438.5$ & $-1.4726$ & $(1,10)$\\
$Pentaquark(1.5GeV)$ & $1.5$ & $0$ & $(2,5)$\\
$K_2(1580)$ & $1575.5$ & $-0.28481$ & $(1,11)$ \\
$K (1630)$ & $1644$ & $0.858896$ & $(8,1)$\\
$K_1 (1650)$ & $1644$ & $-0.36364$ & $(8,1)$\\
$*K^* (1680)$ & $1712.5$ & $(1.934524)0$ & $(1,12)$ \\
$*K_2 (1770)$ & $1781$ & $(0.621469)0$ & $(2,6)$\\ \hline
\end{tabular}
\end{table}

\newpage

\begin{table}
\begin{tabular}{|c|c|c|c|} \hline
Particle and mass & Mass From Formula &  Error \% & $(m,n)$ \\ \hline
$*K^*_3 (1780)$ & $1781$ & $(0.05618)0$ & $(2,6)$\\
$*K_2 (1820)$ & $1849.5$ & $1.620879$ & $(1,13)$ \\
$K (1830)$ & $1849.5$ & $1.065574$ & $(1,13)$\\
$K^*_0 (1950)$ & $1918$ & $-1.64103$ & $(4,2)$\\
$K^*_2 (1980)$ & $1986.5$ & $0.328283$ & $(1,14)$ \\
$*K^*_4 (2045)$ & $2055$ & $(0.488998)0$ & $(2,7)$ \\
$K_2 (2250)$ & $2260.5$ & $0.466667$ & $(1,16)$ \\
$K_3 (2320)$ & $2329$ &  $0.387931$ & $(2,8)$\\
$K^*_5 (2380)$ & $2397.5$ &  $0.735294$ & $(1,17)$\\
$K_4 (2500)$ & $2466$ & $-1.36$ & $(4,4)$ \\
$K (3100)$ & $3082.5$ &  $-0.56452$ & $(1,22)$\\
$*D^\pm (1869.3)$ & $1849.5$ & $-1.05922$ & $(1,13)$  \\
$*D^\pm_0 (1968.5)$ & $1986.5$ & $0.914402$ & $(1,14)$ \\
$*D^*_0 (2007)$ & $1986.5$ & $-1.02143$ & $(1,14)$  \\
$D^*_\pm (2010)$ & $1986.5$ & $-1.16915$ & $(1,14)$ \\
$D_S (2317)$ & $2329$ & $0.51791$ & $(2,8)$ \\
$*D_1 (2420)$ & $2397.5$ & $-0.92975$ & $(1,17)$ \\
$D_1^\pm  (2420)$ & $2397.5$ & $-0.97067$ & $(1,17)$ \\
$D^*_2 (2460)$ & $2466$ & $0.243902$ & $(4,4)$\\
$D^*_\pm (2460)$ & $2466$ &  $0.243902$ & $(4,4)$ \\
$D^\pm_{S1} (2536)$ & $2534.5$ & $-0.07885$ & $(1,18)$  \\
$D_{SJ} (2573)$ & $2534.5$ & $-1.49631$ & $(1,18)$ \\
$*B^{\pm} (5278)$ & $5274.5$ & $ -0.08524$ & $(1,38)$ \\
$*B^0 (5279.4)$ & $5274.5$ & $-0.09281$ & $(1,38)$ \\
$B_j(5732)$ & $5754$ & $-0.47009$ & $(4,10)$ \\
$*B^0_S (5369.6)$ & $5343$ & $-0.49538$ & $(2,19)$ \\
$B^*_{SJ} (5850)$ & $5822.5$ & $-0.47009$ & $(1,42)$ \\
$*B^\pm_c (6400)$ & $6370.5$ & $0.4609$ & $(3,15)$ \\
$*\eta c (1S) (2979)$ & $2945.5$ & $-1.12454$ & $(1,21)$\\
$*J/\psi (1S) (30968)$ & $3082.5$ & $-0.46402$ & $(1,22)$\\
$*\chi c_0 (1P) (3415.1)$ & $3425$ & $0.289889$ & $(2,12)$ \\
$*\chi c_1 (1P) (3510.5)$ & $3493.5$ & $-0.48426$ & $(1,25)$ \\
$*\chi c_2 (1P) (3556)$ & $3562$ & $0.168729$ & $(4,6)$ \\ \hline
\end{tabular}
\end{table}

\newpage

\begin{table}
\begin{tabular}{|c|c|c|c|} \hline
Particle and mass & Mass From Formula &  Error \% & $(m,n)$ \\ \hline
$*\psi (2S) (3685.9)$ & $3699$ &  $0.355408$ & $(2,13)$\\
$*\psi (3770)$ & $3767.5$ & $(-0.06631)0$ & $(1,27)$ \\
$*\psi (3836)$ & $3836$ & $0$ & $(8,3)$\\
$*\psi (4040)$ & $4041.5$ & $(0.037129)0$ & $(1,29)$\\
$*\psi (4160)$ & $4178.5$ &  $(0.444712)0$ & $(1,30)$\\
$*\psi (4415)$ & $4452.5$ & $0.84937$ & $(1,32)$ \\
$*\gamma (1S) (9460.3)$ & $9453$  & $-0.07716$ & $(2,34)$ \\
$\chi b_0 (1P) (9859.9)$ & $9864$ & $0.041583$ & $(16,4)$ \\
$*\chi b_1 (1P) (9892.7)$ & $9864$ & $-0.29011$ & $(16,4)$\\
$*\chi b_2 (1P) (9912.6)$ & $9864$ & $-0.49029$ & $(16,4)$ \\
$*\gamma (2S) (10023)$ & $10001$ & $0.21949$ & $(2,36)$\\
$*\chi b_0 (2P) (10232)$ & $10275$ & $0.42026$ & $(2,37)$ \\
$*\chi b_1 (2P) (10255)$ & $10275$ & $0.1945027$ & $(2,37)$\\
$*\chi b_2 (2P) (10268)$ & $10275$ & $0.068173$ & $(2,37)$\\
$*\gamma (3S) (10355)$ & $10343.5$ & $0.11105$ & $(1,75)$ \\
$*\gamma (4S) (10580)$ & $10549$ & $-0.29301$ & $(2,38)$\\
$*\gamma (10860)$ & $10891.5$ & $0.290055$ & $(3,26)$ \\
$*\gamma (11020)$ & $11028.5$ & $0.077132$ & $(1,80)$ \\ \hline
\end{tabular}
\end{table}

\end{document}